# Optimization of the magnetic properties of aligned Co nanowires / polymer composites

Weiqing Fang, Ioannis Panagiotopoulos, Frédéric Ott and François Boué

*Laboratoire Léon Brillouin CEA/CNRS UMR12, Centre d'Etudes de Saclay, 91191 Gif sur Yvette, France*

Kahina Ait-Atmane and Jean-Yves Piquemal

*ITODYS, Université Paris 7-Denis Diderot, UMR CNRS 7086 2, Place Jussieu,75251 Paris Cedex 05, France*

Guillaume Viau

*Laboratoire de Physique et Chimie des Nano-Objets, INSA de Toulouse, UMR CNRS 5215, 135 Av. de Rangueil, 31077 Toulouse Cedex 4, France*

Florent Dalmas

*Institut de Chimie et des Matériaux Paris-Est (ICMPE), UMR 7182, CNRS/Université Paris-Est Créteil, 2-8 rue Henri Dunant 94320 Thiais, France*

**Abstract**

We aim at combining high coercivity magnetic nanowires in a polymer matrix in a view to fabricate rare-earth free bonded magnets. In particular, our aim is to fabricate anisotropic materials by aligning the wires in the polymer matrix. We have explored the different parameters of the fabrication process in order to produce a material with the best possible magnetic properties. We show that the choice of a proper solvent allows obtaining stable nanowire suspensions. The length and the type of the polymer chains play also an important role. Smaller chains ($M_w$<10000) provide better magnetization results. The magnetic field applied during the casting of the material plays also a role and should be of the order of a fraction of a tesla. The local order of the nanowires in the matrix has been characterized by TEM and Small Angle Neutron Scattering. The correlation between the local order of the wires and the magnetic properties is discussed. Materials with coercivity $\mu_0 H_c$ up to 0.70 T at room temperature have been obtained.

# 1  Introduction

During the last decade, a significant research effort has been put into the fabrication and the investigation of the structure and the magnetic properties of magnetic metal nanowires, especially Ni, Fe and Co (Vazquez 2006) and their alloys (particularly FeNi (Brzòzka et al 1996), CoFe (Qin et al 2002), CoNi (Soumare et al 2008), FePt (Rhen et al 2004). Such wires can be produced either by electrochemical deposition of metals in uniaxial porous templates such as anodic aluminium oxide (Fert et al 1999), by organo-metallic synthesis (Wetz et al 2007), or by the polyol process (Ung et al 2007; Soumare et al 2008 and Soumare et al 2009). Such wires exhibit specific magnetic properties owing to their large shape anisotropy (Sellmyer et al 2001 and McGary et al 2005). The large shape anisotropy of these objects (aspects ratios >5) gives rise to significant coercive fields ($\mu_0 H_c$ = 0.3-1T). The idea of using the shape anisotropy of micro- or nano-objects to produce hard magnetic materials was proposed a long time ago. This property has been used for decades in AlNiCo magnets which consist of FeCo needles in an AlNi matrix (McCurrie 1994). Materials using FeCo elongated single domain particles (Mendelsohn 1955 and Falk 1966) were even commercialized under the name Lodex. Owing to the recent progress in the chemical synthesis techniques we are revisiting this idea.

Cobalt nanowires synthesized by reduction in liquid polyol are mono-crystalline and mono-disperse (Soumare 2009). Their length ranges between 200-300nm and their diameter between 10-20nm. These wires exhibit remarkable hard magnetic properties for a simple 3d metal. Due to the absence of defects and to their mono-crystalline state, experimental coercivity can reach values larger than 0.5 T at room temperature. We showed in previous work that half of the observed coercivity could be accounted for by the magneto-crystalline anisotropy (Maurer 2007 and Ait-Atmane 2013). The remaining part can be attributed to the shape anisotropy contribution.

There are few reports about aligned magnetic nanowires dispersed in a polymer matrix. Fragouli et al (2010) present a technique for magnetic-field-induced formation of magnetic nanowires in a polymer film. They start from a polymer/iron oxide nanoparticle casted solution that is dried under magnetic field. Nanocomposite films with aligned nanowires formed by nanoparticles are obtained. Similar results were obtained by Robbes et al (2011). Park et al (2007) have investigated the micromechanical properties of Ni nanowire/polymer composites.

In this communication we report on the use of a polymer matrix to host aligned Co nanowires in order to fabricate high coercivity anisotropic materials. Among the roles of the different parameters involved, we investigate the role of the polymer matrix (type of polymer, length of the chains), the role of the solvent, and the role of the alignment field. The magnetic bulk properties of the obtained materials were characterized by VSM magnetometry. The microstructures of the materials were investigated by means of MEB, TEM and Small Angle Neutron Scattering. The SANS measurements were performed at the Laboratoire Léon Brillouin on the PAXY spectrometer.

# 2 Sample preparation

## 2.1 Co nanowires synthesis

Co nanowires are synthesized by reduction in liquid polyol (1,2-butanediol) according to procedures previously reported (Soumare 2009). After the synthesis, the magnetic nanowires are separated from the butane-diol synthesis solution by centrifugation and washed 2 times with ethanol and one time with chloroform to remove the remaining organic compounds. At this stage, only a layer of laurate remains at the surface of the nanowires (Ait-Atmane 2013).

## 2.2 Dispersion of the wires in solvent

Several solvents were tested to obtain a dispersion of nanowires. Water has to be strictly excluded since it leads to a fast oxidation of the Co nanowires. Solvents such as toluene, butane-diol and chloroform were tested. In most of the cases, once mixing is done, decantation is complete after a time ranging from minutes to days. In practice, even if the dispersions (~0.5% wt) remain deep black, decantation of the largest magnetic clusters nevertheless takes place. In order to quantify the quality of the dispersions and the ability of a specific solvent to provide a stable suspension, the UV absorbance of different suspensions was measured as a function of time (see figure 1). The behavior is very different for the three solvents investigated, as shown in Figure 1. In the synthesis solution (butane-diol) the evolution of absorbance shows a phase-separation phenomenon. Just after shaking, the solution is dark; in the end, we obtain a supernatant of very low NW concentration butane-diol solution above a far more concentrated liquid. The sharp drop of absorbance after 150 minutes corresponds to the point where the supernatant/dense phase interface passes in front of the optical cell. In the case of toluene solvent, the decantation is very fast: the absorbance decreases from 6 to 1 after only 10 min. On the other extreme, for chloroform, we obtain our best stability results. After a first decantation of the larger particles (over the first 2-3 hours), the suspension stabilizes. In the best cases, suspensions in chloroform have been stable up to several weeks. Note that any incorporation of water leads to an instantaneous decantation.

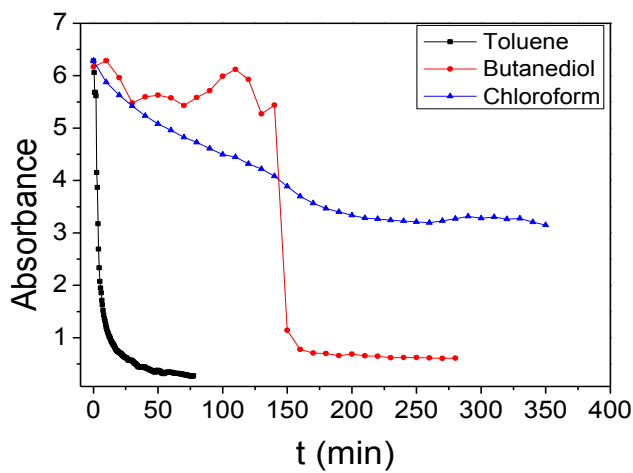

*Figure 1: UV absorbance of cobalt nanowires dispersed in different solvents (1,2-butanediol, toluene, chloroform)* (~0.5% wt) *as a function of time.*

## 2.3 Preparation of solid samples in a polymer matrix

We used polystyrene (PS), Poly(methyl methacrylate) (PMMA) or poly(vinyl pyrrolindone) (PVP) as a matrix for the preparation of the nanocomposites. For PVP, several molecular masses were also considered, namely $M_w$ 10 k, 40 k and 360k. A concentrated polymer chloroform solution (typically 10% v/v polymer let dissolve for several hours) is mixed with a solution of cobalt nanowires which is agitated in an ultrasonic bath during 15 min. The mixture is then mechanically agitated on a shaking table for 30 minutes. The mixture is then poured into a Teflon container and dry cast for 5 days. As the boiling point of chloroform is low (61°C), the container is cooled to about 10°C in order to reduce the evaporation speed of the solvent, which prevents the formation of bubbles on the surface of the nanocomposites film. The composites samples prepared with PS, PMMA and PVP were labeled Co-PS, Co-PMMA and Co-PVP-x, where x stands for the polymer molecular weight.

Moreover, during the preparation of the nanocomposites, the influence of the application of an external magnetic field was evaluated. Two types of experiments were devised: samples prepared by dry casting in a zero magnetic field environment and samples prepared under the application of a magnetic field $H_{align}$ (up to 0.8T) during the casting. Note that this field should be extremely homogeneous with low gradients. Not fulfilling this condition will result in an inhomogeneous sample since the wires will migrate to the extremities of the film during the casting. In order to minimize these gradients we used NMR grade electromagnets with very large polar pieces (200mm in diameter) (see figure 2a). We expect the casting under field to induce an alignment of the wires (see figure 2b), inducing in turn an intrinsic bulk magnetic anisotropy in the sample. This can be readily quantified by magnetometry measurements. The hysteresis cycles can be measured along the direction of the aligned wires (Ox), perpendicular to the wires (Oy) or perpendicular to the sample surface (Oz). Figure 2c shows the principle of expected hysteresis cycles measured along various directions. In the case of an isotropic sample the curve should be rather rounded reflecting the random distribution of the anisotropy directions. In the case of the aligned wires, a measurement with the field applied along the wires long axis should give a rather square hysteresis cycle with a high remanence while a measurement perpendicular to the wires should give a narrow loop with a low coercivity and remanence. We recall that the optimal shape of the hysteresis curve for a permanent magnet is a square one in which the coercivity is at least half of the remanent polarization of the material.

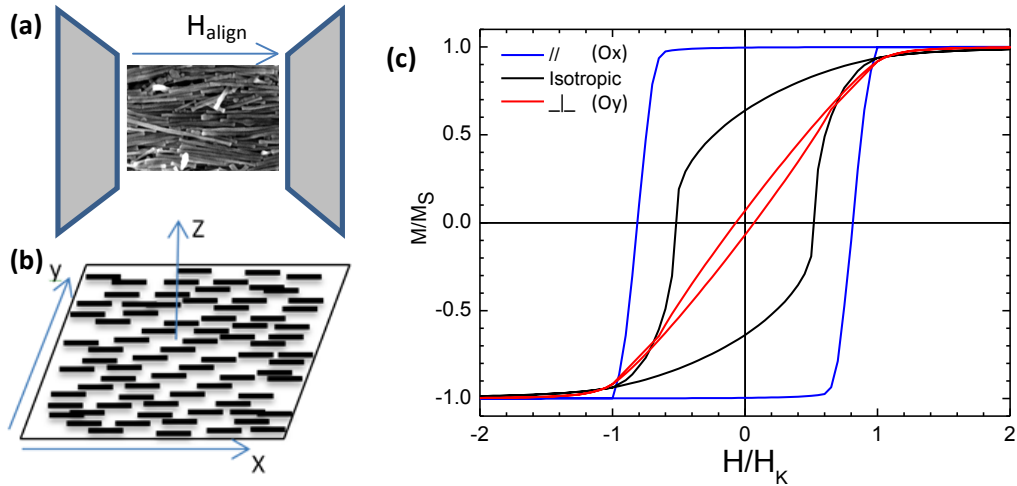

*Figure 2: (a) The wires are aligned between the polar pieces of an electromagnet. (b) Axis convention in an aligned sample; (c) Calculated in-plane hysteresis loops of an isotropic (black) and of an aligned sample (blue, easy axis; red, hard axis).*

# 3 Characterization of the magnetic properties of the composites

## 3.1 Isotropic samples: effect of the polymer matrix on the magnetic properties

Several polymer matrices were tested: polystyrene (PS), poly(methyl methacrylate) (PMMA) and poly(vinyl pyrrolindone) (PVP). The use of PMMA led to very brittle samples. On the contrary, films with PS matrix displayed good mechanical properties. However several tests showed that before casting, the mixture of PVP, chloroform, and NWs was more stable than the equivalent mixture with PS. The mixture with PVP can be stable during 2-3 days while that with PS decant after a few hours. This led us to focus on PVP matrices. The sample is in glassy state. For high molecular weight (e.g. 360 000), the mechanical properties enable an easy handling of the sample. For low molecular weight (<40 000), the samples are quite brittle.

Table 1 shows coercivity and remanence measured on isotropic composites with three polymers of various lengths PS 192 k, PMMA 350 k, and PVP 360 k, 40 k, 10 k. No drastic effect is observed on the coercivity values and the behavior is rather close to a "pure" NW sample in which no polymer has been added (such as the samples deposited on a wafer). Thus we may infer that there has been no significant interaction between the nanowires and the polymer matrix. The normalized remanence values range between $M_r/M_s=0.5$ and $M_r/M_s=2/\pi$ expected for a completely random-3D and in-plane-2D-random distribution of the wire axes respectively. Thus from these values the angular deviation $\Delta\theta$ out of the sample plane (xy-plane, see conventions in Figure 2) can be calculated assuming that the axes distribution is completely random within the sample plane but deviates by $\pm \Delta\theta$ above and below the sample plane. The calculated standard deviation values $\sigma_\theta$ of the out of plane orientation distributions are given in the last row of table 1. It can be noticed that very large chains (especially of PVP360k and PMMA350k) lead to an in-plane 2D-random distribution in contrast to shorter ones - as well as to the "pure" NW sample (with no polymer addition) - which have an intermediate distribution, partially out of plane. When the molecular weight is reduced from 360 k down to 10 k, a measurable increase of the coercivity is observed. This suggests that when small polymer chains are used, they are able to coat and separate the cobalt wires. This can be due to kinetic effects, but here we start from a solution where the different motions should be much faster that the evaporation. So that this can be better associated with "depletion effects" where long chains, due to confinement conformational constraints tend to expel out towards regions free of NW, which turns down to bring the NW together. When the wires are physically separated (by a few nm), the magnetic dipolar interactions are reduced (Maurer 2011). This minimizes the direct effects of dipolar fields between wires and thus increases the effective coercive field.

| Polymer | NW powder on Si | PS 192k | PMMA 350k | PVP 360k | PVP 40k | PVP 10k |
|---|---|---|---|---|---|---|
| $\mu_0 H_c$ (T) | 0.6 | 0.58 | 0.57 | 0.56 | 0.61 | 0.68 |
| $M_r/M_s$ | 0.57 | 0.56 | 0.63 | 0.65 | 0.59 | 0.57 |
| $\sigma_\theta$ (deg) | 27 | 31 | 8 | 0 | 22 | 27 |

*Table 1.   Coercivity and remanence of isotropic composites with three different polymers.*

## 3.2 Effect of the applied field during casting

We studied the effects of the magnitude of the magnetic field applied during the casting of the samples. Two different types of samples were studied (i) cobalt wires dispersed in a chloroform solution without polymer deposited in the form of a thin layer of Co on silicon wafers and (ii) dry cast composites (Co NW + PVP 40k), which is the aim of our study.

The layers of NWs on silicon could be readily characterized by SEM (see Figure 3). It can be observed that the alignment of the wires is invisible for low fields (0.04T). It is barely visible for $\mu_0.H_{align}$ = 0.1 T, but becomes obvious for $\mu_0.H_{align}$=0.2 T and 0.4 T. The results of the magnetometry measurements are summarized in Figure 4 where the coercive field and the remanence have been reported as a function of $H_{align}$. The alignment of the NWs in the polymer matrix can be evaluated by measuring the loop squareness S= $M_R/M_S$ value. For an ideally parallel assembly, a value close to S=1 is expected while it is decreasing to 0.5 for a totally disorganized 3D assembly. The $S_{easy}$=($M_R/M_S$) value that is derived by the measurement along the alignment direction should increase with the alignment while at the same time S=($M_R/M_S$)$_\perp$ value derived by the measurement perpendicular to the field is expected to decrease. The ratio $S_{easy}/S_{hard}$ gives a figure of merit of the anisotropy of magnetic properties obtained by the alignment. It can be noted that even if the SEM image does not show any obvious alignment for $\mu_0.H_{align}$=0.04T, a measurable anisotropy is nevertheless observed ($S_{easy}/S_{hard}$ =1.6). The anisotropy increases almost linearly as a function of $H_{align}$ reaching a ratio $S_{easy}/S_{hard}$ =3 at $\mu_0.H_{align}$ = 0.4T. The coercive fields follow the same trend as a function of the alignment field. The evolution of the coercive field along the easy direction is however not spectacular (increasing only from 0.52 to 0.55T).

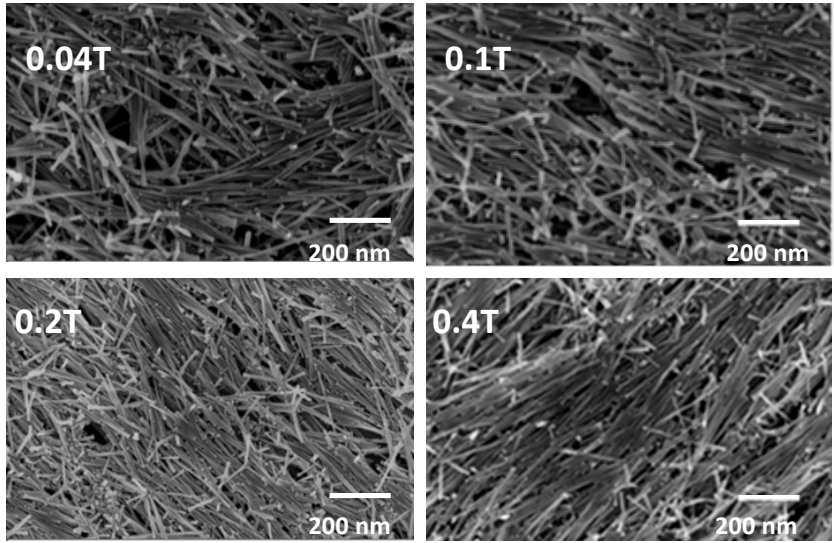

*Figure 3. SEM picture of cobalt nanowires dried on a Si wafer under various magnetic fields.*

The composites films Co-PVP could not be characterized through SEM due to difficulty of preparation. Magnetometry can nevertheless be used to follow the evolution of the anisotropy. The $S_{easy}/S_{hard}$ ratio evolves from 2.2 to 4.25 for an alignment field increasing from 0.2 to 0.8 T. Surprisingly, the coercive field $H_{c,easy}$ decreases from 0.52T down to 0.45T when the alignment field is increased from 0.2 to 0.8T (Fig 4c). In conclusion we can say that alignment fields of a fraction of a tesla should be applied to maximize the anisotropy. However, the level of anisotropy is not directly correlated with the coercive field. It is likely that the applied field magnitude modifies the clustering of the wires. Under larger drying fields, larger NW clusters are formed. This increases the dipolar interactions and consequently reduces the effective coercive field. In the case of the wires deposited on a Si substrate the drying process is faster and the formation of larger clusters is limited.

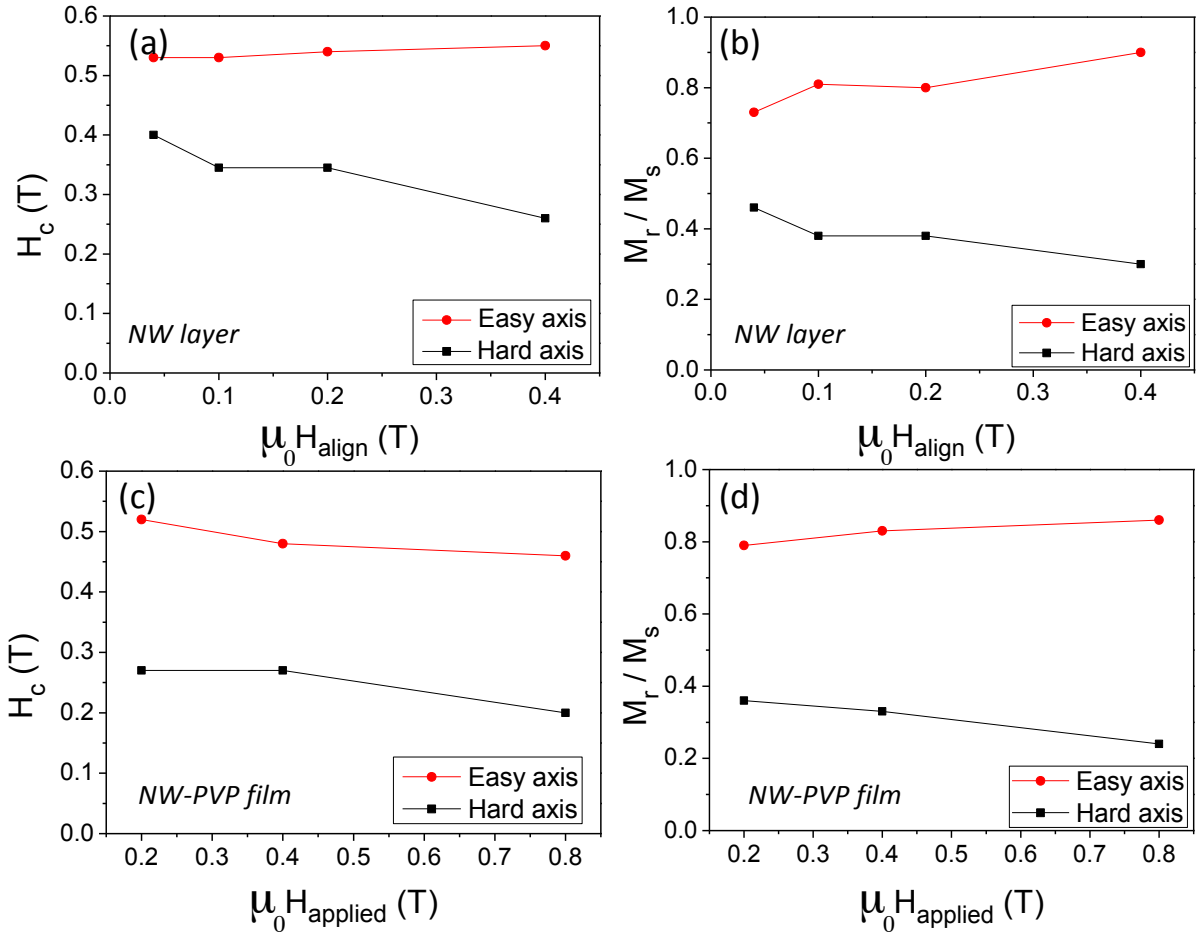

*Figure 4: Coercivity and remanence of the cobalt nanowires dispersed on a Si substrate (a,b) and in PVP 40k casting under magnetic field (c,d).*

# 4 Characterization of the morphology and the dispersion of cobalt nanowires

## 4.1 Microscopic and macroscopic observations

In order to correlate more finely the local arrangement of wires with the magnetic properties, we performed systematic studies on 2 cast samples (Co-PS-192 and Co-PVP-360). TEM, Small Angle Neutron Scattering and magnetometry were performed on these samples. Figure 5 shows TEM images of ultramicrotomic slices of the (Co-PS) sample prepared without (Fig. 5a) or under (Fig. 5b) the application of an external magnetic field. In the case of the absence of an external magnetic field during drying, isotropic samples are obtained and one observes small bundles of 6-12 aligned wires (size of order 500 nm - 1 μm) together with more open clusters of NW independent in orientation, with similar size. In the case of the aligned samples, one can see that the wires are well aligned with an angular dispersion of 18° FWHM (as deduced from an analysis in the *imageJ* software). The wires are still grouped in small bundles of 10-20 wires. The Co-PVP-360 sample preparation for the TEM could not be achieved with our microtome procedure because the sample is hydrophilic, which is not compatible with the standard preparation method.

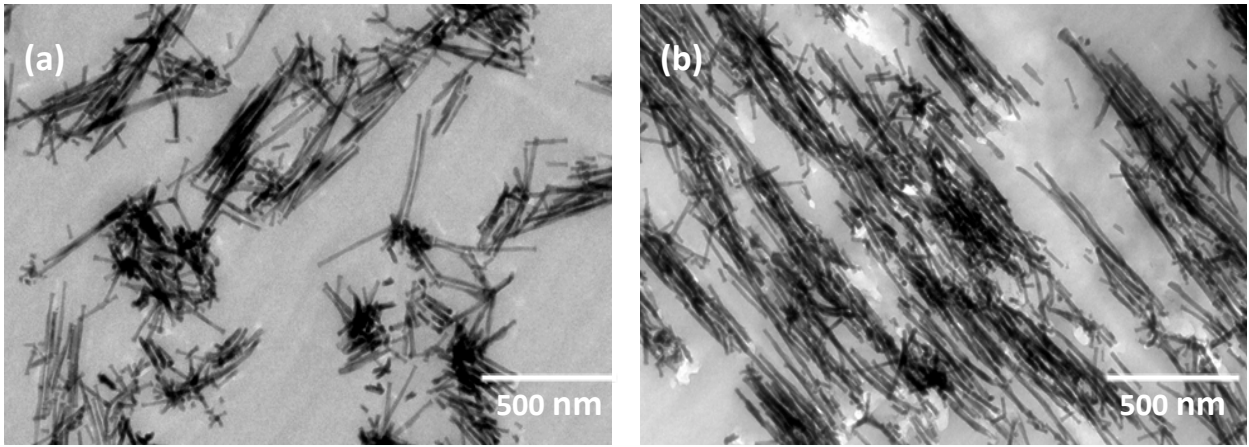

*Figure 5. TEM pictures of cobalt nanowires dispersed in PS 192k and dried without (a) and (b) with the application of an external magnetic field ($\mu_0H$ = 0.6 T).*

The magnetometry characterization is shown in Figure 6. Looking at the values in detail shows that the Co-PS sample exhibits slightly better properties than the Co-PVP sample in terms of coercive field. For the Co-PS sample, we obtain $\mu_0H_c$ =0.70 T, while for the Co-PVP sample $\mu_0H_c$ = 0.65 T. On the other hand, the PVP sample is better aligned since for Co-PS, $M_r/M_s$ = 0.9 while for the Co-PVP sample $M_r/M_s$ = 0.93. This is confirmed by SANS scattering data (Fig. 7). The alignment is more marked in the PVP sample.

In order to get a more quantitative interpretation of the magnetization data we have fitted the loops using the Stoner-Wohlfarth (SW) model for an assembly of particles with the angular distributions suggested by the remanence values. More specifically we have assumed the following distribution of angular orientations: (i) distribution of the azimuthal angle ɸ between the wire axis and the alignment field direction within the sample plane and (ii) distribution of the angle θ between the wire axis and sample plane. Note that θ as defined here is the complementary of the polar angle typically used in spherical coordinates. Gaussian distributions with zero average values and standard deviations $\sigma_\phi$, $\sigma_\theta$ respectively have been assumed. The results are shown as the continuous lines in Fig.6. The agreement is fair considering that the SW model is based on the hypothesis of reversal by homogeneous magnetization rotation of non-interacting magnetic entities, which may be violated in our case mainly by dipolar interactions (Maurer, 2011). Consequently the deviations of the fitted curves are maximal in the demagnetization quadrant of the loops where interaction assisted nucleation events are more likely to occur. The PS sample has identical $\sigma_\phi$, $\sigma_\theta$ values of 20°. Note that these values are consistent with the direct space observations. The PVP is better aligned within the plane ($\sigma_\phi$=14°) but the out of plane value is similar. A further indication of departure from the ideal SW behavior is that the best value of anisotropy field required to describe the easy axis curves ($H_K$(par)) is about 2/3 of that of the hard axis curve $H_K$(per). In principle these two values should coincide.

Theoretically for a wire of hexagonal Co with an aspect ratio of 20 and its crystallographic c-axis along its length, a value of $H_K$=2.3 T is expected, of which 1.56 T due to shape anisotropy and the rest 0.74 T due to the magneto-crystalline anisotropy. This theoretical value is closer to the ones obtained by the perpendicular measurement. This is due to the fact that in the field range close to saturation, homogeneous reversible rotation mechanisms are dominant whereas in the range near the magnetization switching point more complex mechanisms and irreversible jumps may occur. We have used the value that best describes the former range for the hard axis loop fit. Furthermore in the curves measured with the applied field perpendicular to the easy direction (hard axis), there is an apparent contradiction between the anisotropy value which must be used to account for the high field range (approach to saturation) as opposed to the value that predicts the correct loop shape. In conclusion the real anisotropy field is best represented by $H_K(per)$ while the deviation of the ratio $H_K(par)/H_K(per)$ from unity gives a measure of the dominance of non-homogeneous reversal mechanisms. In that sense the PVP sample which is better aligned shows a larger deviation from the ideal SW-behavior and despite its more marked anisotropy has a lower coercivity. An explanation can be that alignment allows the wires to become more closely packed, which increases dipolar interactions. Progress in orientation and separation can thus be contradictory in some situations.

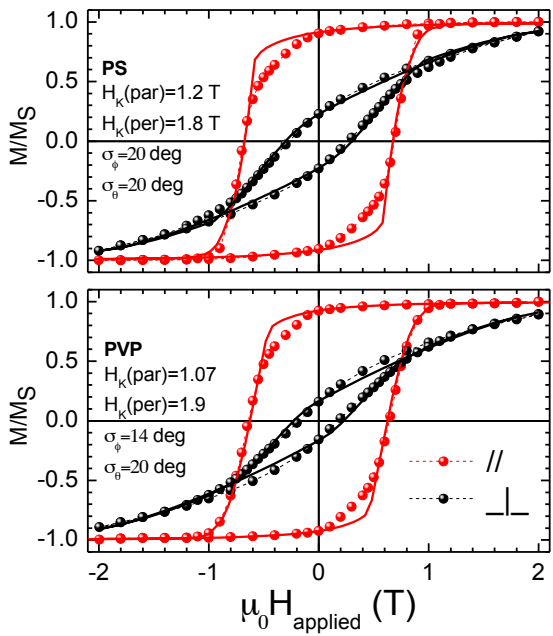

*Figure 6. Hysteresis loops of cobalt nanowires dispersed in PS 192k and PVP 360k, aligned at room temperature.*

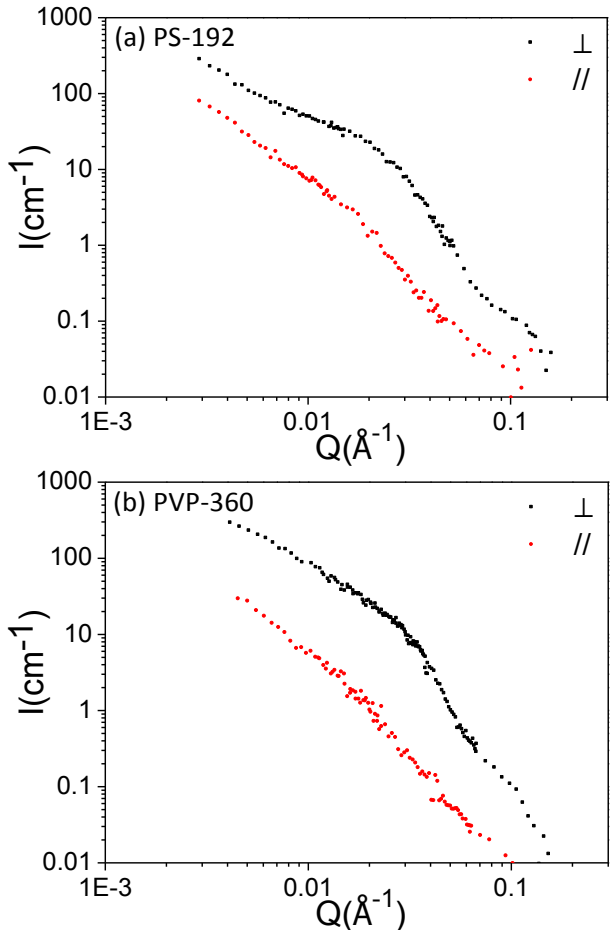

*Figure 7: SANS scattering for (a) the Co-PS-192 sample and (b) the Co-PVP-360 sample. The scattered intensity is measured parallel and perpendicular to the average wires direction. The anisotropy as measured by the ratio of the signal // and ⊥ is far more marked in the PVP sample.*

## 4.2 Discussion

Macroscopic observations show that the NW dispersed in (PVP + chloroform) decant slower than dispersion in (PS + chloroform). A natural conclusion is that the nanowires are better dispersed in the (PVP-chloroform) solution, and are present in the solution as smaller clusters. During casting, when starting from a more dispersed suspension it is expected that a better local alignment is obtained. This is confirmed by SANS data (Fig. 7) which show a better degree of orientation of the nanowires at the nanometric scale. It is confirmed by bulk magnetic measurements in which one observes that the ratio $S_{easy}/S_{hard}$ is higher in PVP compared to PS composites. However, the better orientation of the wires does not lead to overall better magnetic properties since the coercive field of the PVP sample is smaller than in the PS sample. This can be interpreted by the fact that the wires in the PVP sample are better aligned but are also in closer contact. This leads to an overall loss of effective coercivity due to dipolar interactions. On the other hand, when the system is more disordered, the wires are physically more separated and thus the dipolar interactions are reduced. Thus in order to optimize the properties of these composites, a subtle balance between order and magnetic properties must be found.

# 5 Conclusions

We have quantified the effects of various parameters on the fabrication of Co nanowires/polymer composites for the fabrication of permanent magnets. In order to optimize the magnetic properties, magnetic fields as large as possible should be applied. The field gradients must however be minimized to obtain an homogeneous material. Small molecular weight of the polymer chains enable to obtain better macroscopic magnetic properties by leading to a better separation of the nanowires (less aggregation, better dispersion). This minimizes the local magnetic dipolar interactions between individual wires and provides better bulk magnetic properties. In the case of anisotropic materials, being able to produce highly aligned materials does not lead to optimal magnetic properties especially in terms of coercive field. This can also be accounted for by the increased magnetic dipolar interactions between wires which are in close contact when perfectly aligned. One has thus to find an optimum between the quality of the dispersion, the degree of alignment and the bulk magnetic properties. Our experimental observations are presently being modeled using micromagnetic simulations to figure out the exact role of the local ordering between wires on the magnetic properties. The best achieved coercivity was 0.70T, with a remanence $M_r/M_s$ of 0.9. Assuming a Co volume fraction of 60%, the maximal energy product $(BH)_{max}$ of such a material would be on the order of 260 kJ/m$^3$ which is in the range of the best SmCo magnets ($(BH)_{max}$~ 120-200 kJ/m$^3$). This is not yet competitive with the best NdFeB magnets ($(BH)_{max}$~ 440kJ/m$^3$). However such compounds might present interest in high temperature applications (above 250°C) (Ait-Atmane, 2013) where NdFeB magnets cannot be used unless very expensive Dy is added as for example in automotive applications.

This work was partially supported by the EU-NMP funded project "Rare-Earth Free Permanent Magnets" REFREEPERMAG.